\documentclass{tlp}

\usepackage{latexsym}

\newcommand\bcmdtab{\noindent\bgroup\tabcolsep=0pt%
  \begin{tabular}{@{}p{10pc}@{}p{20pc}@{}}}
\newcommand\ecmdtab{\end{tabular}\egroup}

\newcommand{\At}{\mbox{{\it At}}}
\newcommand{\SSM}{\mbox{{\it SSM}}}
\newcommand{\LSM}{\mbox{{\it LSM}}}
\newcommand{\vph}{\varphi}
\newcommand{\n}{\mbox{\bf{not}}}

\newtheorem{theorem}{Theorem}[section]
\newtheorem{lemma}[theorem]{Lemma}

  \title[Theory and Practice of Logic Programming]
        {Computing large and small stable models\footnote{This is a
         full version of an extended abstract presented at the 
         International Conference on Logic Programming, ICLP-99 and 
         included in the proceedings published by MIT Press.}}

  \author[M. Truszczy\'nski]
         {Miros\l aw Truszczy\'nski\\
         Department of Computer Science, University of Kentucky,
         Lexington, KY 40506-0046, USA\\
         \email{mirek@cs.uky.edu}}

\pagerange{\pageref{firstpage}--\pageref{lastpage}}

\begin{document}

\label{firstpage}

\maketitle

\begin{abstract}
In this paper, we focus on the problem of existence and computing
of {\em small} and {\em large} stable models. We show that for every
fixed integer $k$, there is a linear-time algorithm to decide the
problem $\LSM$ (large stable models problem): does a logic program $P$
have a stable model
of size at least $|P|-k$? In contrast, we show that the problem $\SSM$
(small stable models problem) to decide whether a logic program $P$ has
a stable model of size at most $k$ is much harder. We present two
algorithms for this problem but their running time is given by
polynomials of order depending on $k$. We show that the problem $\SSM$
is {\em fixed-parameter intractable} by demonstrating that it is
$W[2]$-hard. This result implies that it is unlikely an algorithm
exists to compute stable models of size at most $k$ that would run in
time $O(m^c)$, where $m$ is the size of the program and $c$ is a constant
independent of $k$. We also provide an upper bound on the
fixed-parameter
complexity of the problem $\SSM$ by showing that it belongs to the
class $W[3]$.
\end{abstract}

\section{Introduction}

The stable model semantics by Gelfond and Lifschitz \cite{gl88} is one 
of the two most widely studied semantics for normal logic programs, 
the other one being the well-founded semantics by Van Gelder, Ross and 
Schlipf \cite{vrs91}. Among 2-valued semantics, the stable model
semantics is commonly regarded as the one providing the correct meaning to 
the negation operator in logic programming. It coincides with the least 
model semantics on the class of Horn programs, and with the well-founded 
semantics and the perfect model semantics on the class of stratified 
programs \cite{abw87}. In addition, the stable model semantics is closely 
related to the notion of a default extension by Reiter \cite{mt89c,bf91a}.
Logic programming with stable model semantics has applications in knowledge
representation, planning and reasoning about action. It was also recently 
proposed as a computational paradigm well suited for solving 
combinatorial optimization and constraint satisfaction problems 
\cite{mt99,nie99}.

Before we proceed, we will recall the definition of a stable model of a
logic program, and some related terminology and properties. The
reader is referred to \cite{mt93} for a more detailed treatment of the
subject. In the paper we deal only with the propositional case. A logic 
program {\em rule} is an expression $r$ of the form
\[
r=\ \ \ a \leftarrow b_1,\ldots,b_s,\n(c_1),\ldots,\n(c_t),
\]
where $a$, $b_i$s and $c_i$s are propositional atoms. The atom $a$ is
called the {\em head} of $r$ and is denoted by $h(r)$. Atoms $b_i$ and
$c_i$ form the {\em body} of $r$. The set $\{b_1,\ldots,b_s\}$ is called the
{\em positive} body of $r$ (denoted by $b^+(r)$) and the set
$\{c_1,\ldots,c_t\}$ is called the {\em negative} body of $r$ (denoted
by $b^-(r)$). A {\em logic program} is a collection of rules. For a
logic program $P$, by $\At(P)$ we denote the set of atoms occurring in
its rules and by $h(P)$ --- the set of atoms appearing as the heads
of rules in $P$. We will also denote the {\em size} of $P$, that is,
the total number of occurrences of atoms in $P$, by $size(P)$. Throughout 
the paper we use $n$ to denote the number of atoms in a logic program $P$, 
and $m$ to denote the size of $P$.

A set of atoms $M \subseteq \At(P)$ {\em satisfies} a rule $r$ if 
$h(r)\in M$, or if $b^+(r)\setminus M\not=\emptyset$, or if $b^-(r)\cap
M\not=\emptyset$. A set of atoms $M\subseteq \At(P)$ is a {\em model} of a
program $P$ if $M$ satisfies all rules of $P$.  

A logic program rule $r$ is called {\em Horn} if $b^-(r)=\emptyset$. A
{\em Horn program} is a program whose every rule is a Horn rule.
The intersection of two models of a Horn program $P$ is a model of $P$. 
Since the set of all atoms is a model of $P$, it follows that every Horn
program $P$ has a unique least model. We will denote this model by
$LM(P)$. The least model of a Horn program $P$ can be constructed by
means of the van Emden-Kowalski operator $T_P$ \cite{vEK76}.
Given a Horn program $P$ and a set of atoms $M\subseteq P$, we define
\[
T_P(M)=\{a\colon a\leftarrow b_1,\ldots,b_s\in P,\ \mbox{and}\
\{b_1,\ldots,b_s\}\subseteq M\}.
\]
We also define
\[
T^0_P(M) =\emptyset, \ \ \mbox{and}\ \ \ T^{i+1}_P(M) = T_P(T^i_P(M)).
\]
Since the operator $T_P$ is monotone, the sequence $T^{i}_P(\emptyset)$
is monotone and its union yields the least model of a Horn program $P$.
That is, 
\[
LM(P)=\bigcup_{i=0}^\infty T^i_P(\emptyset).
\]
If $P$ is finite, the sequence stabilizes after finitely many steps.

For a logic program rule $r$, by $horn(r)$ we denote the rule obtained
from $r$ by eliminating all negated atoms from the body of $r$. If $P$
is a logic program, we define $horn(P)=\{horn(r)\colon r\in P\}$.

Let $P$ be a logic program (possibly with rules containing negated
atoms). For a set of atoms $M\subseteq \At(P)$ we define the {\em
reduct} of $P$ with respect to $M$ to be the program obtained by 
eliminating from $P$ each rule
$r$ such that $b^-(r)\cap M\not=\emptyset$ (we call such rules {\em
blocked} by $M$), and by removing negated atoms from all other rules in
$P$. The resulting program is a Horn program. We will denote it
by $P^M$. As a Horn program, $P^M$ has the least model $LM(P^M)$. If 
$M=LM(P^M)$, $M$ is a {\em stable} model of $P$. Clearly, if $M$ is a
stable model of $P$, $M\subseteq h(P)$. Both the notion of the
reduct and of a stable model are due to Gelfond and Lifschitz
\cite{gl88}. 

In the paper we restrict our attention to programs whose rules do not
contain multiple positive occurrences of the same atom nor
multiple negative occurrences of the same atom in the body. It is clear
that adopting this assumption does not limit the generality of our
considerations. Repetitive occurrences can be eliminated in linear time 
(in the size of the program) and doing so does not affect stable
models of the program.

If $M$ is a stable model of $P$, each rule $r$ such that
$b^+(r)\subseteq M$ and $b^-(r)\cap M=\emptyset$ (that is, such that $M$
satisfies its body), is called a {\em generating} rule for $M$. Clearly, 
if $M$ is a stable model of $P$, it is also a stable model of the program 
consisting of all rules in $P$ that are generating for $M$. 

There are several ways to look at the search space of possible stable
models of a program $P$. The most direct way is to look for stable models 
by considering all candidate subsets of $h(P)$. For each candidate subset
$M\subseteq h(P)$, one can compute the corresponding reduct $P^M$, its
least model $LM(P^M)$, and check the equality $M=LM(P^M)$ to decide
whether $M$ is stable. An alternative way is to observe that stable
models are determined by subsets of the set of atoms appearing negated 
in $P$. Indeed, let us denote this set by $Neg(P)$ and let us consider sets
$M\subseteq \At(P)$ and $B\subseteq Neg(P)$. Let $B'=Neg(P)\setminus B$. 
Then, $M$ is a stable model of $P$ if and only if $M=LM(P^{B'})$, 
$B\cap M=\emptyset$ and $B'\subseteq M$. Thus, the existence of stable 
models can be decided by considering subsets of $Neg(P)$. Finally, 
one can consider the search 
space of all subsets of $P$ itself, and regard each such subset as a 
candidate for the set of generating rules of a stable model. Indeed, if
$M\subseteq \At(P)$ and $P'\subseteq P$, then $M$ is a stable model of 
$P$ if and only if $M=h(P')$, $P'$ is the set of all generating rules 
for $M$ in $P$ and $M=LM(horn(P'))$. 

The problem with the stable model semantics is that, even in 
the propositional case, reasoning with logic programs under the stable 
model semantics is computationally hard. It is well-known that deciding 
whether a finite 
propositional logic program has a stable model is NP-complete \cite{mt88}. 
Consequently, it is not at all clear that logic programming with the stable 
model semantics can serve as a practical computational tool.
 
This issue can be resolved by implementing systems computing stable 
models and by experimentally studying the performance of these systems. 
Several such projects are now under way. Niemel\"a and Simons \cite{ns96} 
developed a system, {\em smodels}, for computing stable models of finite
function symbol-free logic programs and reported very promising 
performance results. For some classes of programs, {\em smodels} 
decides the existence of a stable model in a matter of seconds even if 
an input program consists of tens of thousands of clauses. Encouraging 
results on using 
{\em smodels} to solve planning problems are reported in \cite{nie99}. 
Another well-advanced system is DeReS \cite{cmt96}, designed to compute 
extensions of arbitrary propositional default theories but being especially 
effective for default theories encoding propositional logic programs. 
Finally, systems capable of reasoning with disjunctive logic programs were 
described in \cite{elmps97} and \cite{adn97}.

However, faster implementations will ultimately depend on better
understanding of the algorithmic aspects of reasoning with logic programs 
under the stable model semantics. In this paper, we investigate the complexity 
of deciding whether a finite propositional logic program has stable models 
of some restricted sizes. Specifically, we study the following 
two problems ($|P|$ stands for the number of rules in a logic program 
$P$):
\begin{description}
\item[$\LSM$] (Large stable models) Given a finite propositional logic 
program $P$ and an integer $k$, decide whether there is a stable model 
of $P$ of size at least $|P|-k$.
\item[$\SSM$] (Small stable models) Given a finite propositional logic 
program $P$ and an integer $k$, decide whether there is a stable model 
of $P$ of size no more than $k$.
\end{description}

Inputs to the problems $\LSM$ and $\SSM$ are pairs $(P,k)$, where $P$ is
a finite propositional logic program and $k$ is a non-negative integer. 
Problems of this type are referred to as {\em parametrized} decision
problems. By fixing a parameter, a parameterized decision problem gives
rise to its {\em fixed-parameter} version. In the case of problems $\LSM$
and $\SSM$, by fixing $k$ we obtain the following two fixed-parameter
problems ($k$ is now no longer a part of input): 
\begin{description}
\item[$\LSM(k)$] Given a finite propositional
logic program $P$, decide whether $P$ has a stable model of size at
least $|P| -k$.
\item[$\SSM(k)$] Given a finite propositional
logic program $P$, decide whether $P$ has a stable model of size at most
$k$.
\end{description}

The problems $\LSM$ and $\SSM$ are NP-complete. It follows
directly from the NP-completeness of the problem of existence of stable 
models \cite{mt88}. But fixing $k$ makes a difference!
Clearly, the fixed-parameter problems $\SSM(k)$ and $\LSM(k)$ can be solved 
in polynomial time (unlike the problems $\SSM$ and $\LSM$ which, most likely, 
cannot). Indeed, consider a finite propositional logic program $P$. 
Then, there are $O(n^k)$ subsets of $\At(P)$ (in fact, as pointed out
earlier, it is enough to consider subsets of $h(P)$ or $Neg(P)$) of cardinality at most $k$
(we recall that in the paper $n$ stands for the number of atoms in $P$). 
For each such subset $M$, it can be checked in time linear in $m$ ---
the size of $P$ --- whether $M$ is a stable model of $P$. Thus, 
one can decide whether $P$ has a stable model of size at most $k$ in time 
$O(mn^k)$. 

Similarly, there are only $O(|P|^k)$ subsets of $P$ of size at
least $|P|-k$. Each such subset is a candidate for the set of 
generating rules of a stable model of size at least $|P|-k$ 
(and smaller subsets, clearly, are not). Given such a subset $R$, 
one can check in time $O(m)$ whether $R$ generates a stable model 
for $P$. Thus, it follows that there is an algorithm that decides in 
time $O(m|P|^k)$ whether a logic program $P$ has 
a stable model of size at least $|P|-k$. 

While both algorithms are polynomial in the size of the program, their
asymptotic complexity is expressed by the product of the size of a
program and a polynomial of order $k$ in the number of atoms of the
program or in the number of rules of the program. Even for small values 
of $k$, say for $k\geq 4$, the functions $mn^k$ and $m|P|^k$ grow very 
fast with $m=size(P)$, $n=|\At(P)|$ and $|P|$, and render the 
corresponding algorithms infeasible.

An important question is whether algorithms for problems $\SSM(k)$ and
$\LSM(k)$ exist whose order is significantly lower than $k$, preferably,
a constant independent of $k$. The study of this question is the
main goal of our paper. A general framework for such investigations was 
proposed by Downey and Fellows \cite{df97}. They introduced the concepts of 
{\em fixed-parameter tractability} and {\em fixed-parameter
intractability} that are defined in terms of a certain
hierarchy of complexity classes known as the $W$ {\em hierarchy}. 

In the paper, we show that the problem $\LSM$ is fixed-parameter tractable 
and demonstrate an algorithm that for every fixed $k$ decides the problem 
$\LSM(k)$ in linear time --- a significant improvement over the
straightforward algorithm presented earlier. 

On the other hand, we demonstrate that the problem $\SSM$ is much harder. 
We present an algorithm to decide the problems $\SSM(k)$, for $k\geq 1$,
that is asymptotically faster than the simple algorithm described above 
but the improvement is rather insignificant. Our algorithm runs in time 
$O(mn^{k-1})$, an improvement only by the factor of $n$.
The difficulty in finding a substantially better algorithm is
not coincidental. We provide evidence that the
problem $\SSM$ is {\em fixed-parameter intractable}. This result implies it 
is unlikely that there is an algorithm to decide the problems
$\SSM(k)$ whose running time would be given by a polynomial of order 
independent of $k$. 

The study of fixed-parameter tractability of problems occurring in the
area of nonmonotonic reasoning is a relatively new research topic. Another
paper that pursues this direction is \cite{gss99}. The authors focus
there on parameters describing structural properties of programs and
show that in some cases, fixing these parameters leads to polynomial
algorithms. 

Our paper is organized as follows. In Section \ref{ifp}, we recall basic
concepts of the theory of fixed-parameter intractability by Downey and
Fellows \cite{df97}. The following two sections present the algorithms
to decide the problems $\LSM $ and $\SSM$, respectively. The next section
focuses on the issue of fixed-parameter intractability of the problem 
$\SSM$ and contains the two main results of the paper. The last section 
contains conclusions and open problems.

\section{Fixed-parameter intractability}\label{ifp}

This section recalls basic ideas of the work of Downey and Fellows 
on fixed-parameter intractability. The reader is referred to \cite{df97} 
for a detailed treatment of this subject.

Informally, a {\em parametrized} decision problem is a decision problem
whose inputs are pairs of items, one of which is referred to as a {\em
parameter}. The graph colorability problem is an example of a parametrized 
problem. The inputs are pairs $(G,k)$, where $G$ is an undirected graph and 
$k$ is a non-negative integer. The problem is to decide whether $G$ can 
be colored with at most $k$ colors. Another example is the vertex cover 
problem in a graph. Again, the inputs are graph-integer pairs $(G,k)$ and 
the question is whether $G$ has a vertex cover of cardinality $k$ or 
less. The problems $\SSM$ and $\LSM$ are also examples of parametrized 
decision problems.
Formally, a {\em parametrized} decision problem is a set
$L\subseteq \Sigma^*\times\Sigma^*$, where $\Sigma$ is a fixed
alphabet. 

By selecting a concrete value $\alpha\in \Sigma^*$ of the parameter, a
parametrized decision problem $L$ gives rise to an associated 
{\em fixed-parameter} problem $L_\alpha =\{x:(x,\alpha)\in L\}$.
For instance, by fixing the value of $k$ to 3, we get a fixed-parameter
version of the colorability problem, known as 3-colorability. Inputs to
the 3-colorability problem are graphs and the question is to decide
whether an input graph can be colored with 3 colors. Clearly, the
problems $\SSM(k)$ ($\LSM(k)$, respectively) are fixed-parameter versions 
of the problem $\SSM$ ($\LSM$, respectively).

The interest in the fixed-parameter problems stems from the
fact that they are often computationally easier than the corresponding
parametrized problems. For instance, the problems $\SSM$ and
$\LSM$ are NP-complete yet, as we saw earlier, their parametrized versions 
$\SSM(k)$ and $\LSM(k)$ can be solved in polynomial time. Similarly, 
the vertex cover problem is NP-complete but its fixed-parameter versions 
are in the class P. To see this, observe that to decide whether a graph
has a vertex cover of size at most $k$, where $k$ is a fixed value and
not a part of an input, it is enough to generate all subsets with at
most $k$ elements of the vertex set of a graph, and then check if any of 
them is a vertex cover. A word of caution is in order here. It is not 
always the case that fixed-parameter problems are easier. For instance, 
the 3-colorability problem is still NP-complete.

As we already pointed out, the fact that a problem admits a polynomial-time 
solution does not necessarily mean that practical algorithms to solve it
exist. An algorithm that runs in time $O(N^{15})$, where $N$ is the
size of the input, is hardly more practical than an algorithm with an
exponential running time (and may even be a worse choice in practice).
The algorithms we presented so far to argue that the problems $\SSM(k)$,
$\LSM(k)$ and the fixed-parameter versions of the vertex cover problem 
are in P rely on searching through the space of $N^k$ possible
solutions (where $N$ is the number of atoms of a program, the number of
rules of a program, or the number of vertices in a graph, respectively). 
Thus, these
algorithms are not practical, except for the very smallest values of $k$.
The key question is how fast those polynomial-time solvable fixed-parameter 
problems can really be solved. Or, in other words, can one significantly
improve over the brute-force approach?  

A technique to deal with such questions is provided by the fixed-parameter
intractability theory of Downey and Fellows \cite{df97}. A parametrized
problem $L\subseteq \Sigma^*\times\Sigma^*$ is {\em fixed-parameter 
tractable} if there exist a constant $p$, an integer 
function $f$ and an algorithm $A$ such that $A$ determines whether 
$(x,y)\in L$ in time $f(|y|)|x|^p$ ($|z|$ stands for the length of 
a string $z\in \Sigma^*$). The class of fixed-parameter tractable
problems will be denoted by FPT. Clearly, if a parametrized problem $L$
is in FPT, each of the associated fixed-parameter problems $L_y$ is 
solvable in polynomial time by an algorithm whose exponent does not depend 
on the value of the parameter $y$. It is known (see \cite{df97}) that the
vertex cover problem is in FPT. 

There is substantial evidence to support a conjecture that
some parametrized problems whose fixed-parameter versions are in P are
not fixed-parameter tractable. To study and compare complexity of 
parametrized problems Downey and Fellows proposed the following notion 
of reducibility\footnote{The definition given here is sufficient for the
needs of this paper. To obtain  structural theorems a subtler definition 
is needed. This topic goes beyond the scope of the present paper. The
reader is referred to \cite{df97} for more details.}. A parametrized 
problem $L$ can be {\em reduced} to a parametrized problem $L'$ if there 
exist a constant $p$, an integer function $q$ and an algorithm $A$ that 
to each instance $(x,y)$ of $L$ assigns an instance $(x',y')$ of $L'$ 
such that
\begin{enumerate}
\item $x'$ depends upon $x$ and $y$ and $y'$ depends upon $y$ only,
\item $A$ runs in time $O(q(|y|)|x|^p)$,
\item $(x,y)\in L$ if and only if $(x',y')\in L'$.
\end{enumerate}
Downey and Fellows also defined a hierarchy of complexity classes called
the {\em W hierarchy}:
\begin{equation}\label{eq1}
{\rm FPT} \subseteq {\rm W[1]} \subseteq {\rm W[2]} \subseteq {\rm W[3]}
\ldots
\end{equation}

The classes W[t] can be described in terms of problems that are complete
for them (a problem $D$ is {\em complete} for a complexity class $\cal
E$ if $D\in{\cal E}$ and every problem in this class can be reduced to 
$D$). Let us call a boolean formula {\em $t$-normalized} if it is of the 
form of product-of-sums-of-products ... of literals, with $t$ being 
the number of products-of, sums-of expressions in this definition.
For example, 2-normalized formulas are products of sums of literals.
Thus, the class of 2-normalized formulas is precisely the class of
CNF formulas. We define the {\em weighted $t$-normalized
satisfiability problem} as:
\begin{description}
\item[$WS(t)$] Given a $t$-normalized formula $\vph$, decide whether
there is a model of $\vph$ with exactly $k$ atoms (or, alternatively, 
decide whether there is a satisfying valuation for $\vph$ which assigns 
the logical value {\bf true} to exactly $k$ atoms)
\end{description}
Downey and Fellows show that for $t\geq 2$, the problems 
$WS(t)$ are complete for the class W[t]. They also show that 
a restricted version of the problem $WS(2)$:
\begin{description}
\item[$WS_3(2)$] Given a 3CNF formula $\vph$ and an integer $k$
(parameter), decide whether there is a model of $\vph$ with exactly $k$
atoms
\end{description} 
is complete for the class $W[1]$. Downey and Fellows conjecture that all 
the implications in (\ref{eq1}) are proper\footnote{If true, this conjecture
would imply that in the context of fixed-parameter tractability there is 
a difference between the complexity of weighted satisfiability for 3CNF and 
CNF formulas.}. In particular, they conjecture that problems in the classes
W[t], with $t\geq 1$, are not fixed-parameter tractable.

In the paper, we relate the problem $\SSM$ to the problems $WS(2)$ and
$WS(3)$ to place the problem $\SSM$ in the W hierarchy, to obtain 
estimates of its complexity and to argue for its fixed-parameter 
intractability.

\section{Large stable models}

In this section we will show an algorithm for the parametrized problem 
$\LSM$ that runs in time $O(2^{k+k^2}m)$, where $(P,k)$
is an input instance and, as in all other places in the paper, $m=size(P)$. 
This result implies that the problem $\LSM$ is
fixed-parameter tractable and that there is an algorithm that for every 
fixed $k$ solves the problem $\LSM(k)$ in linear-time. 
 
Given a logic program $P$, denote by $P^*$ the logic program 
obtained from $P$ by eliminating from the bodies of the rules in $P$ all
literals $\n(a)$, where $a$ is not the head of any rule from $P$.
The following well-known result states the key property of the program 
$P^*$.

\begin{lemma}\label{th-10}
A set of atoms $M$ is a stable model of a logic program $P$ if and only
if $M$ is a stable model of $P^*$.
\end{lemma}

Lemma \ref{th-10} implies that the problem $\LSM$ 
has a positive answer for $(P,k)$ if and only if it has a positive answer for 
$(P^*,k)$. Moreover, it is easy to see that $P^*$ can be constructed from 
$P$ in time linear in the size of $P$. Thus, when looking for algorithms
to decide the problem $\LSM$ we may restrict our attention to programs 
$P$ in which every atom appearing negated in the body of a rule appears 
also as the head of a rule (that is, to such programs $P$ for which we
have $Neg(P)\subseteq h(P)$). 

By $P^k$ let us denote the program consisting of those rules $r$ in 
$P$ for which $|b^-(r)|\leq k$. We have 
the following lemma. 

\begin{lemma}\label{lem-42}
Let $P$ be a logic program such that $Neg(P)\subseteq h(P)$. Let 
$M\subseteq \At(P)$ be a set of atoms such that $|M|\geq |P|-k$. Then: 
\begin{enumerate}
\item $M$ is a stable model of $P$ if and only if $M$ is a stable model of 
$P^k$
\item if $M$ is a stable model of $P^k$, then $P^k$ has no more than 
$k+k^2$ different negated literals appearing in the bodies of
its rules.
\end{enumerate}
\end{lemma}
Proof: (1) Consider a rule $r\in P\setminus P^k$. Then $|b^-(r)|\geq k+1$ 
and, consequently, $b^-(r)\cap M\not=\emptyset$. Indeed, if $b^-(r)\cap M 
=\emptyset$, then $|M\cup b^-(r)| = |M|+|b^-(r)| > |P|$. 
Since $Neg(P)\subseteq h(P)$, $b^-(r)\subseteq h(P)$. In addition, 
(both if we assume that $M$ is a stable model of $P$ and if we assume 
that $M$ is a
stable model of $P^k$), we have $M\subseteq h(P)$. Thus,
$b^-(r)\cup M\subseteq h(P)$. Now observe that
$|P|\geq |h(P)|$. Thus, $|M\cup b^-(r)|\leq |h(P)| \leq |P|$, a contradiction.

Since for every rule $r\in P\setminus P^k$ we have $b^-(r)\cap
M\not=\emptyset$, it follows that $(P^k)^M = P^M$.
Hence, $M = LM(P^M)$ if and only if $M=LM((P^k)^M)$. Consequently, $M$ is a
stable model of $P$ if and only if $M$ is a stable model of $P^k$. 

\noindent
(2) Let $P'$ be the set of rules from $P^k$ such that $r\in P'$
if and only if $b^-(r)\cap M=\emptyset$ (the rules in $P'$ contribute to
the reduct $(P^k)^M$) and let $P''$ be the set of the
remaining rules in $P^k$ (these are the rules that are
eliminated when the reduct $(P^k)^M$ is computed). Since
$Neg(P)\subseteq h(P)$,
for every rule $r\in P$, $b^-(r)\subseteq h(P)$. Thus,
$\bigcup \{b^-(r)\colon r\in P'\} \subseteq h(P)\setminus M$.
Since $M\subseteq h(P)$ (as $M$ is a stable model of $P^k$) and 
$|P|\geq |h(P)|$, we have $|\bigcup \{b^-(r)\colon r\in P'\}|\leq k$.
Further, since $|P'|\geq |M|\geq |P|-k\geq |P^k|-k$, it follows 
that $|P''|\leq k$. Consequently, 
$|\bigcup\{b^-(r)\colon r\in P''\}|\leq k^2$.
Hence, the second part of the assertion follows. \hfill $\Box$

Let us now consider the following algorithm for the problem $\LSM(k)$ 
(the input to this algorithm is a logic program $P$).
\begin{enumerate}
\item Eliminate from the input logic program $P$ all literals $\n(a)$, 
where $a$ is not the head of any rule from $P$. Denote the resulting
program by $Q$.
\item Compute the set of rules $Q^k$ consisting of those rules $r$ in
$Q$ for which $|b^-(r)|\leq k$.
\item Decide whether $Q^k$ has a stable model $M$ such that
$|M|\geq |Q|-k$.
\end{enumerate}

This algorithm reports YES if and only if the program $Q^k$ has a stable 
model $M$ such that $|M|\geq |Q|-k$. By Lemma \ref{lem-42}, that happens
precisely if and only if $Q$ has a stable model $M$ such that
$|M|\geq |Q|-k$. This last statement, by Lemma \ref{th-10}, is equivalent 
to the statement that $P$ has a stable model $M$ such that $|M|\geq
|P|-k$. In other words, our algorithm correctly decides the problem
$\LSM(k)$. 

Let us notice that steps 1 and 2 can be implemented in time
$O(m)$, where the constant hidden by the ``big O'' notation does
not depend on $k$. To implement step 3, let us recall that every stable 
model of a logic program is determined by some subset of the set of
atoms that appear negated in the program (each such subset uniquely
determines the reduct, as we stated in the introduction; see also
\cite{btk93}). By Lemma \ref{lem-42}, the set of such atoms in the 
program $Q^k$ has cardinality at most ${k+k^2}$. Checking for each 
subset of this set whether it determines a stable model of $Q^k$ can be
implemented in time $O(size(Q^k))=O(m)$. Consequently,
our algorithm runs in time $O(2^{k+k^2}m)$ (with the
constant hidden by the ``big O'' notation independent of $k$).

\begin{theorem}
The problem $\LSM$ is fixed-parameter tractable. Moreover, for each fixed 
$k$ there is a linear-time algorithm to decide whether a logic program 
$P$ has a stable model of size at least $|P|-k$.
\end{theorem} 

\section{Computing stable models of size at most $k$}

In the introduction we pointed out that there is a straightforward 
algorithm to decide the problem $\SSM(k)$ that runs in 
time $O(m n^{k})$, where $m=size(P)$ and $n=|\At(P)|$. For $k\geq 1$
(the assumption we adopt in this section), this algorithm can be slightly 
improved. Namely, we will now describe an algorithm for the problem 
$\SSM(k)$ that runs in time $O(F(k)m n^{k-1})$, where $F$ is some 
integer function. Thus, if $k$ is fixed and not a part of the input, 
this improved algorithm runs in time $O(mn^{k-1})$. 

We present our algorithm under the assumption that input logic programs 
are {\em proper}. We say that a logic program rule $r$ is {\em proper} 
if:
\begin{description}
\item[(P1)] $h(r)\notin b^+(r)$, and
\item[(P2)] $b^+(r)\cap b^-(r)=\emptyset$
\end{description}
We say that a logic program $P$ is {\em proper} if all its rules are
proper. Rules that violate at least one of the conditions (P1) and (P2)
(that is, rules that are not proper) have no influence on the collection
of stable models of a program as we have the following well-known result
(see, for instance, \cite{bradix96jlp1}).

\begin{lemma}\label{l-11}
A set of atoms $M$ is a stable model of a logic program $P$ if and only
if $M$ is a stable model of the subprogram of $P$ consisting of all
proper rules in $P$.
\end{lemma}

It is easy to see that rules that violate (P1) or (P2) can be
eliminated from a logic program $P$ in time $O(m)$. Thus,
the restriction to proper programs does not affect the generality of
our discussion.

For a proper logic program $P$ and for a set $A\subseteq \At(P)$ of
atoms, we define $P(A)$ to be the program consisting of all those 
rules $r$ of $P$ that are not blocked by $A$ (in other words, those 
that satisfy $b^-(r)\cap A=\emptyset$) and whose positive body is 
contained in $A$ (in other words, such that $b^+(r)\subseteq A$). 

Let $P$ be a logic program and let $A\subseteq \At(P)$ be a set
of atoms. A stable model $M$ of $P$ is called {\em $A$-based} if
\begin{enumerate}
\item $M$ is of the form $A\cup\{a\}$, where $a\in \At(P)\setminus A$,
and
\item $M\subseteq LM(P(A)^M)$ (in other words, when computing $LM(P^M)$, 
the derivation of $A$ does not require that $a$ be derived first).
\end{enumerate}

We have the following simple lemma.
\begin{lemma}\label{aa}
Let $k$ be an integer such that $k\geq 1$. A proper logic program $P$ 
has a stable model of cardinality $k$ if and only if for some $A\subseteq 
\At(P)$, with $|A|=k-1$, $P$ has an $A$-based stable model. 
\end{lemma}

It follows from Lemma \ref{aa} that when deciding the existence of
$k$-element stable models, $k\geq 1$, it is enough to focus on the 
existence of $A$-based stable models.  
This is the approach we take here. In most general terms, our algorithm 
for the problem $\SSM(k)$ consists of generating all subsets 
$A \subseteq \At(P)$, with $|A|\leq k-1$, and for each such subset $A$, 
of checking whether $P$ has an $A$-based stable model. This latter task 
is the key.  

We will now describe an algorithm that, given a logic program $P$ and 
a set $A\subseteq \At(P)$, decides whether $P$ has an 
$A$-based stable model. To this end, we define $P'(A)$ to be the
program consisting of all those rules $r$ of $P$ such that:
\begin{enumerate}
\item $b^-(r)\cap A=\emptyset$ ($r$ is not blocked by $A$)
\item $h(r)\notin A$
\item $b^+(r)\setminus A$ consists of exactly one element; {\em we will
denote it by $a_r$}.
\end{enumerate}
Our algorithm is based on the following result
allowing us to restrict attention to the program $P(A)$ 
(the statement of the lemma and its proof rely on the terminology 
introduced above).

\begin{lemma} \label{l-12}
Let $A$ be a set of atoms. A proper logic program $P$ has an $A$-based
stable model if and only if $P(A)$ has an $A$-based stable model
$M=A\cup \{a\}$, such that $a\notin \{a_r\colon r\in P'(A)\}$.
\end{lemma}
Proof: ($\Rightarrow$) Let $M$ be an $A$-based stable model of $P$.
Assume that $M=A\cup \{a\}$, for some $a\notin A$. Since $P(A)^M
\subseteq P^M$, $LM(P(A)^M)\subseteq LM(P^M) = M$. Since $M$ is
$A$-based, we have that  $M\subseteq LM(P(A)^M)$. It follows that $M$ 
is an $A$-based stable model of $P(A)$. 

Let us assume that there is a rule $s\in P'(A)$ such that $a=a_s$.
The rule $s$ is not blocked by $A$. Since $a\in b^+(s)$, we have that
$a\notin b^-(s)$ (we recall that all rules in $P$ are proper). Hence, 
$s$ is not blocked by $\{a\}$ either. Consequently, $horn(s)\in P^M$.
Since $s\in P'(A)$, the body of $horn(s)$ (that is, $b^+(s)$) is 
contained in $M$. The set
$M$ is a least model of $P^M$. In particular, $M$ satisfies $horn(s)$. 
Thus, it follows that $h(s)\in M$. In the same time, $h(s)\not= a$
(as $s$ is proper). Thus, $h(s)\in A$, a contradiction (we recall 
that $s\in P'(A)$). It follows that $a\notin \{a_r\colon r\in P'(A)\}$.

\noindent
($\Leftarrow$) We will now assume that $M=A\cup \{a\}$ is an $A$-based
stable model of $P(A)$ such that $a\notin \{a_r\colon r\in P'(A)\}$.
Similarly as before, we have $M=LM(P(A)^M)\subseteq LM(P^M)$. Let us
assume that $LM(P^M)\setminus M\not=\emptyset$. Then there is a rule 
$t$ in $P^M$ such that the body of $t$ is contained in $M$ and $h(t)
\notin M$. Let $s$ be a rule in $P$ that gives rise to $t$ when
constructing the reduct. Assume first that the body of $t$ (that is,
$b^+(s)$) is contained in $A$. Then $s\in P(A)$, $t\in P(A)^M$ and, 
consequently, $h(t)\in LM(P(A)^M)=M$, a contradiction. 

Thus, the body of $t$ is not contained in $A$. Since the body of $t$
is contained in $M$, it consists of $a$ and, possibly, some other 
elements, all of which are in $A$. It follows that 
$s\in P'(A)$. Consequently, $a=a_s$ and $a\in \{a_r\colon r\in P'(A)\}$,
a contradiction. Thus, $LM(P^M)=M$, that is, $M$ is a stable model of $P$.
Since $M=LM(P(A)^M)$, it follows that $M$ is an $A$-based model of $P$.
\hfill $\Box$

Let $A$ be a set of atoms. A logic program with negation, $P$, is an
{\em $A$-program} if $P=P(A)$, that is if for every rule $r\in P$ we
have $b^+(P)\subseteq A$ and $b^-(P)\cap A = \emptyset$.
Clearly, the program $P(A)$, described above, is an $A$-program. We will
now focus on $A$-programs and their $A$-based stable models.

Let $A$ be a set of atoms. We denote by $R(A)$ the set of all proper
Horn rules over the set of atoms $A$. Clearly, the cardinality of 
$R(A)$ depends on the cardinality of 
$A$ only.
Further, we define ${\cal P}(A)$ to be the set of all Horn 
programs $Q\subseteq R(A)$ satisfying the condition $LM(Q)=A$. 
As in the case of $R(A)$, the cardinality of ${\cal P}(A)$ also 
depends on the size of $A$ only.

We will now describe conditions that determine whether an $A$-program 
$P$ has an $A$-based stable model. To this end, with every atom $a\in 
\At(P)\setminus A$, we associate the following values:
\begin{itemize}
\item $F(a)= 1$ if there is a rule $s$ in $P$ with $h(s)\notin A\cup\{a\}$ 
and $a\notin b^-(s)$; $F(a) =0$, otherwise
\item $G(a)=\ $ the number of rules $s$ in $P$ with $h(s)=a$ and $a\notin
b^-(s)$. 
\end{itemize}
Further, with every proper Horn rule $r\in R(A)$ and every atom $a\in 
\At(P)\setminus A$, we associate the quantity:
\begin{itemize}
\item $H(r,a)= 1$ if there is a rule $s$ in $P$ with $horn(s)=r$ and 
$a\notin b^-(s)$; $H(r,a)=0$, otherwise.
\end{itemize}

The following lemma characterizes $A$-based stable models of an
$A$-program. Both the statement of the lemma and its proof rely on the
terminology introduced above.

\begin{lemma} \label{l-main}
Let $A$ be a set of atoms, let $P$ be an $A$-program and let $a$ be an
atom such that $a\in \At(P)\setminus A$. Then $A\cup\{a\}$ is an 
$A$-based stable model of $P$ if and only if $F(a) = 0$, $G(a) > 0$, and 
for some program $Q\in {\cal P}(A)$ and for every rule $r\in Q$, 
$H(r,a) > 0$.
\end{lemma}
Proof: $(\Rightarrow)$ We denote $M = A\cup \{a\}$ and assume that $M$
is an $A$-based stable model for $P$. It follows that $M = LM(P^M)$.
Let $P_A$ be the subprogram of $P$ consisting of those rules of $P$ whose
head belongs to $A$. Since $M$ is an $A$-based stable model of $P$, we 
have $A = LM(P_A^M)$. Let $Q$ be the
program obtained from $P_A^M$ by removing multiple occurrences of 
rules. Clearly, $Q\in {\cal P}(A)$. It follows directly from the 
definition of the reduct that for every rule $r\in Q$, $H(r,a) = 1$.

Next, we observe that $a\in LM(P^M)$. Thus, $G(a) > 0$. Let us assume
that $F(a) = 1$. Let $r$ be a rule in $P$ such that $h(r)\notin
A\cup\{a\}$ and $a\notin b^-(r)$. Since $P$ is an $A$-program, $A\cap 
b^-(r)=\emptyset$. Thus, it follows that $horn(r)\in P^M$. We also have 
that $b^+(r)\subseteq A\subseteq M$. Since $M$ is a model of $P^M$,
$h(r)\in M$. However, in the same time we have that $h(r) \notin A\cup\{a\} 
(=M)$, a contradiction. It follows that $F(a)=0$.

\noindent
$(\Leftarrow)$ We now assume that for some $a\in \At(P)\setminus A$,
$F(a) = 0$, $G(a) > 0$ and for some program $Q\in {\cal P}(A)$ and for 
every rule $r\in Q$, $H(r,a) = 1$. As before, we set $M=A\cup \{a\}$. We
will show that $M = LM(P^M)$.
 
First, since $P$ is an $A$-program and $H(r,a)=1$ for every rule $r\in 
Q$, it follows that $Q\subseteq P(A)^M$. Thus, $A\subseteq LM(P(A)^M)$. Second, 
we have that $G(a)> 0$. Thus, there is a rule $r\in P$ such that $h(r)=a$ 
and $a\notin b^-(r)$. It follows that $horn(r)\notin Q$ and $horn(r)\in P^M$.  
Since $Q\subseteq P(A)^M$, $A= LM(Q)$ and $b^+(r)\subseteq A$, we obtain
that $a\in
LM(P(A)^M)$. Thus, $M\subseteq LM(P(A)^M)$.
Finally, since $F(a)=0$, we have that for every rule $s\in P$ such that 
$a\notin b^-(s)$, $h(s)\in M$. Thus, $LM(P^M)$ does not contain any 
atom not in $M$. Consequently, $M= LM(P^M)$ and $M$ is a stable model of
$P$. Since $M\subseteq LM(P(A)^M)$, $M$ is an $A$-based stable model of
$P$. \hfill$\Box$

We will discuss now effective ways to compute values $F(a)$, $G(a)$ and 
$H(r,a)$. Clearly, computing the values $G(a)$ can be accomplished in
time linear in the size of the program, that is, in time $O(m)$. Indeed,
we start by initializing all values $G(a)$ to 0. Then, for each rule
$s\in P$, we set $G(h(s)):=G(h(s))+1$ if $h(s)\notin b^-(s)$, and leave
$G(h(s))$ unchanged, otherwise. To decide which is the case requires
that we scan all negated lierals in the body of $s$. That takes time
$O(|b^-(s)|)$. Thus, the overall time is $O(m)$.

Computing values $F(a)$ and $H(r,a)$ is more complicated. First, we prove 
the following lemma.

\begin{lemma}\label{vn}
Let $P$ be an $A$-program, let $a\in \At(P)\setminus A$ and let $r\in
R(A)$. Then
\begin{enumerate}
\item $F(a)=1$ if and only if $a\notin \bigcap \{\{h(s)\}\cup b^-(s):
s\in P, h(s)\notin A\}$.
\item $H(r,a)=1$ if and only if $a \notin \bigcap \{b^-(s): s\in P,
horn(s)=r\}$.
\end{enumerate}
\end{lemma}
Proof: (1) Let us assume first that $F(a)=1$. Then there is a rule $s\in
P$ such that $h(s)\notin A\cup \{a\}$ and $a\notin b^-(s)$. Thus,
$a\notin \{h(s)\}\cup b^-(s)$. Consequently, the identity 
$a\notin \bigcap \{\{h(s)\}\cup b^-(s):s\in P, h(s)\notin A\}$ follows. All
the implications in this argument can be reversed. Hence, we obtain
the assertion (1). 

\noindent
(2) Let us assume that $H(r,a)=1$. Then, there is a rule $s\in P$ such
that $horn(s)=r$ and $a\notin b^-(s)$. Consequently, $a \notin \bigcap 
\{b^-(s): s\in P, horn(s)=r\}$. As in (1), all the implications are in fact
equivalences and the assertion (2) follows. \hfill $\Box$

Lemma \ref{vn} shows that to compute all the values $F(a)$ one
has to compute the set 
\[
\bigcap \{\{h(s)\}\cup b^-(s):s\in P, h(s)\notin A\}.
\]
To this end, for each atom $a$ we will compute the number of sets in
$\{\{h(s)\}\cup b^-(s):s\in P, h(s)\notin A\}$ that $a$ is a member of.
We will denote this number by $C(a)$. We first initialize all values
$C(a)$ to 0. Then, we consider all sets in $\{\{h(s)\}\cup b^-(s):s\in P,
h(s)\notin A\}$ in turn. For each such set and for each atom $a$ in this
set we set $C(a):=C(a)+1$. The set $\bigcap \{\{h(s)\}\cup
b^-(s):s\in P, h(s)\notin A\}$ is given by all those atoms $a$ for which
$C(a)$ is equal to the number of sets in $\{\{h(s)\}\cup b^-(s):s\in
P, h(s)\notin A\}$. It is clear that the time needed for this
computation is linear in the size of the program (assuming appropriate
linked-list representation of rules). Thus, all the values $F(a)$ can be
computed in time linear in the size of the program, that is, in $O(m)$
steps.

To compute values $H(r,a)$ we proceed similarly. First, we compute all
the sets $\{s:s\in P, horn(s)=r\}$, where $r\in R(A)$. To this end, we
scan all rules in $P$ in order and for each of them we find the rule $r\in
R(A)$ such that $horn(s)=r$. Then we include $s$ in the set $\{s:s\in P,
horn(s)=r\}$. Given $s$, it takes $O(g|A|)$ steps to identify rule $r$
(where $g$ is some function). Indeed, the size of $b^+(s)$ is bound by
$|A|$ as $P$ is an $A$-program. Moreover, the number of rules in $R(A)$
depends on $|A|$ only. Thus, the task of computing all sets $\{s:s\in P,
horn(s)=r\}$, for $r\in R(A)$, can be accomplished in $O(g(|A|)|P|)$ 
steps. Next, for each these sets of rules, we proceed as in the case of 
values $F(a)$, to compute their intersections. Each such computation takes 
time $O(m)$, where $m=size(P)$). Thus, computing all the values $H(r,a)$ 
can be accomplished in time $O(g(|A|)|P|+|R(A)|m) = O(f(|A|)m)$, for 
some function $f$.

We can now put all the pieces together. As a result of our
considerations, we obtain the following algorithm for
deciding the problem $\SSM(k)$. 

\begin{tabbing}
\quad\=\quad\=\quad\=\quad\=\quad\=\quad\=\quad\=\quad\=\quad\=\quad\=\quad\=\\
{\bf Algorithm to decide the problem $\SSM(k)$, $k\geq 1$}\\
{\bf Input:} A logic program $P$ ($k$ is {\em not} a part of input)\\
\ \\
(0)\>\>\>{\bf}if $\emptyset$ is a stable model of $P$ {\bf then} return YES
and exit;\\
(1)\>\>\>$P:=$ the set of proper rules in $P$;\\
(2)\>\>\>{\bf for} every $A\subseteq \At(P)$ with $|A|\leq k-1$ {\bf do}\\
(3)\>\>\>\>compute the set of rules $R(A)$ and the set of programs 
         ${\cal P}(A)$;\\
(4)\>\>\>\>compute the program $P(A)$;\\
(5)\>\>\>\>compute the program $P'(A)$ and the set $B=\{a_r\colon r\in
P'(A)\}$;\\
(6)\>\>\>\>given $P(A)$ and $R(A)$, compute tables $F$, $G$ and $H$ (as
             described above);\\ 
(7)\>\>\>\>{\bf for} every $a\in \At(P(A))\setminus A\setminus B$ {\bf do}\\
(8)\>\>\>\>\>{\bf if}\\
(9)\>\>\>\>\>\>$F(a) = 0$, $G(a) > 0$ {\bf and}\\
(10)\>\>\>\>\>\>there is a program $Q\in {\cal P}(A)$ s. t.
             for every rule $r\in Q$, $H(r,a) > 0$\\
(11)\>\>\>\>\>{\bf then} report YES and exit;\\
(12)\>\>\>report NO and exit.
\end{tabbing}

The correctness of this algorithm follows from Lemmas \ref{aa} -
\ref{l-main}. We will now analyze the
running time of this algorithm. Clearly, line (0) can be executed in
$O(m)$ steps. As we already observed,
rules that are not proper can be eliminated from $P$ in time $O(m)$.
Next, there are $O(n^{k-1})$ iterations of loop (2). In each of them,
line (3) takes time $O(f_1(k))$, for some function $f_1$ (let us recall 
that $|R(A)|$ and $|{\cal P}(A)|$ depend on $|A|$ only). Further, 
lines (4) and (5) can be executed in time $O(m)$. 
Line (6), as we discussed earlier, can be implemented so that to run in
$O(f(k)m)$ steps. Loop (7) is executed $O(n)$ times and each iteration takes
$O(f_2(k))$ steps, for some function $f_2$ (let us again recall 
that $|{\cal P}(A)|$ depends on $k$ only). Thus, the running time of
the whole algorithm is $O(F(k)mn^{k-1})$, for some integer
function $F$.
Consequently, we get the following result.

\begin{theorem}
There is an integer function $F$ and an algorithm ${\cal A}$ such that
${\cal A}$ decides the problem $\SSM(k)$ and runs in time
$O(F(k)mn^{k-1})$ (the constant
hidden in the "big Oh" notation does not depend on $k$).
\end{theorem}

\section{Complexity of the  problem $\SSM$}\label{ssm}

The algorithm outlined in the previous section is not quite satisfactory. 
Its running time is still high. A natural question to ask is: are there 
significantly better algorithms for the problems $\SSM(k)$?
In this section we address this question by studying the complexity 
of the problem $\SSM$. Our goal is to show that the problem is difficult 
in the sense of the W hierarchy. We will show that the problem $\SSM$ is 
$W[2]$-hard and that it is in the class W[3]. To this end,
we define the {\em $(\leq k)$-weighted $t$-normalized satisfiability 
problem} as:
\begin{description}
\item[$WS^\leq(t)$] Given a $t$-normalized formula $\vph$, decide whether
there is a model of $\vph$ with at most $k$ atoms ($k$ is a parameter). 
\end{description}
The problem $WS^\leq(t)$ is a slight variation of the problem $WS(t)$.
It is known to be complete for the class W[t], for $t\geq 2$ (see
\cite{df97}, 
page 468). To show W[2]-hardness of $\SSM$, we will reduce 
the problem $WS^\leq(2)$ to the problem $\SSM$. Given the overwhelming
evidence of fixed-parameter intractability of problems that are
$W[2]$-hard \cite{df97}, it is unlikely that algorithms for problems 
$\SSM(k)$ exist whose asymptotic behavior would be given by a polynomial 
of order independent of $k$. To better delineate the location of the
problem $\SSM$ in the W hierarchy we also provide an upper bound on
its hardness by showing that it can be reduced to the problem 
$WS^\leq(3)$, thus proving that the problem $\SSM$ belongs to the class $W[3]$.

We will start by showing that the problem $\SSM(k)$ is reducible (in the
sense of the definition from Section \ref{ifp}) to the problem $WS^\leq(3)$. 
To this end, we describe an encoding of a logic program $P$ by means of a 
collection of clauses $T(P)$ so that $P$ has a stable model of size at most 
$k$ if and only if $T(P)$ has a model with no more than $(k+1)(k^2+2k)$ 
atoms. In the general setting of the
class NP, an explicit encoding of the problem of existence of stable
models in terms of propositional satisfiability was described in
\cite{bed94}. Our encoding, while different in key details, uses
some ideas from that paper.

Let us consider an integer $k$ and a logic program $P$. For each
atom $q$ in $P$ let us introduce new atoms $c(q)$, $c(q,i)$, $1\leq
i\leq k+1$, and $c^-(q,i)$, $2\leq i\leq k+1$. Intuitively, atom $c(q)$ 
represents the fact that in the process of computing the
least model of the reduct of $P$ with respect to some set of atoms, atom
$q$ is computed no later than during the iteration $k+1$ of the van
Emden-Kowalski operator. Similarly, atom $c(q,i)$ 
represents the fact that in the same process atom $q$
is computed exactly in the iteration $i$ of the van Emden-Kowalski 
operator. Finally, atom $c^-(q,i)$, expresses the fact that 
$q$ is computed {\em before} the iteration $i$ of the van Emden-Kowalski
operator. The formulas $F_1(q,i)$, $2\leq i\leq k+1$, and $F_2(q)$ describe 
some basic relationships between atoms $c(q)$, $c(q,i)$ and $c^-(q,i)$
that we will require to hold: 
\[
F_1(q,i) = c^-(q,i) \Leftrightarrow c(q,1)\vee\ldots\vee c(q,i-1),
\]
\[
F_2(q) = c(q) \Leftrightarrow c(q,1)\vee\ldots\vee c(q,k+1).
\]
 
Let $r$ be a rule in $P$ with $h(r)=q$, say 
\[
r = q\leftarrow a_1,\ldots,a_s,\n(b_1),\ldots,\n(b_t).
\]
We define a formula $F_3(r,i)$, $2\leq i\leq k+1$, by
\[
F_3(r,i) = c^-(a_1,i)\wedge\ldots\wedge c^-(a_s,i)\wedge
       \neg c(b_1)\wedge \ldots\wedge \neg c(b_t) \wedge \neg c^-(q,i).
\]
We define $F_3(r,1)= {\bf false}$ ({\bf false} is a distinguished
contradictory formula in our propositional language) if $s\geq 1$. 
Otherwise, we define 
\[
F_3(r,1) = \neg c(b_1)\wedge \ldots\wedge \neg c(b_t).
\]
Speaking informally, formula $F_3(r,i)$ asserts that $q$ is 
computed by means of rule $r$ in the iteration $i$ of the least model 
computation process and that it has not been computed earlier.

Let $r_1,\ldots,r_v$ be all rules in $P$ with atom $q$ in the head.
We define a formula $F_4(q,i)$, $1\leq i\leq k+1$, by
\[
F_4(q,i) = c(q,i) \Leftrightarrow F_3(r_1,i)\vee\ldots \vee F_3(r_v,i).
\]
Intuitively, the formula $F_4(q,i)$ asserts that when computing the
least model of the reduct of $P$, atom $q$ is first computed in the
iteration $i$.

We now define the theory $T_0(P)$ that encodes the problem of existence
of small stable models:
\begin{eqnarray*}
T_0(P)&=& \{F_1(q,i)\colon q\in \At(P),\ 2\leq i\leq k+1\}\cup
      \{F_2(q)\colon q\in \At(P)\}\cup\\
      & & \{F_4(q,i)\colon q\in \At(P),\ 1\leq i\leq k+1\}. 
\end{eqnarray*}

Next, we establish some useful properties of the theory $T_0(P)$.
First, we consider a set $U$ of atoms that is a model of $T_0(P)$ and define
\[
M(U)=\{q\in \At(P)\colon c(q)\in U\}.
\]
\begin{lemma}\label{lem-11}
Let $U$ be a model of $T_0(P)$ and let $q\in M(U)$. Then there is a unique 
integer $i$, $1\leq i\leq k+1$, such that $c(q,i)\in U$.
\end{lemma}
Proof: Since $U$ is a model of a formula $F_2(q)$, there is an 
integer $i$, $1\leq i\leq k+1$, such that $c(q,i)\in U$. To prove
uniqueness of such $i$, assume that there are
two integers $j_1$ and $j_2$, $1\leq j_1<j_2\leq k+1$, such that
$c(q,j_1)\in U$ and $c(q,j_2)\in U$. Since $U\models 
F_4(q,j_2)$, it follows that there is a rule $r\in P$ with $h(r)=q$
and such that $U\models F_3(r,j_2)$. In particular,
$U\models \neg c^-(q,j_2)$. In the same time, since
$c(q,j_1)\in U$ and $U\models F_1(q,j_2)$, we have
$c^-(q,j_2)\in U$, a contradiction. \hfill $\Box$

For every atom $q\in M(U)$ define $i_q$ to be the integer whose existence
and uniqueness is guaranteed by Lemma \ref{lem-11}. Define 
$i_U=\max\{i_q\colon q\in M(U)\}$. Next, for each $i$, $1\leq i\leq
i_U$, define
\[
[M(U)]_i=\{q\in M(U)\colon i_q=i\}.
\]

\begin{lemma}\label{lem-12}
Let $U$ be a model of $T_0(P)$. Under the terminology introduced above,
for every $i$, $1\leq i\leq i_U$, $[M(U)]_i\not=\emptyset$. 
\end{lemma}
Proof: We will proceed by downward induction. By the definition of 
$i_U$, $[M(U)]_{i_U}\not=\emptyset$. Consider $i$, $2\leq i \leq i_U$, 
and assume
that $[M(U)]_i\not =\emptyset$. We will show that $[M(U)]_{i-1}\not 
=\emptyset$. 
Let $q\in [M(U)]_i$. Clearly, $c(q,i)\in U$ and, since $U\models F_4(q,i)$, 
there is a rule $r = q\leftarrow a_1,\ldots,a_s,\n(b_1),\ldots,\n(b_t)$ 
such that $U\models F_3(r,i)$. Consequently, for every $j$,
$1\leq j\leq s$, $c^-(a_j,i)\in U$. Assume that for every $j$,
$1\leq j\leq s$, $c^-(a_j,i-1)\in U$. Since $U\models c^-(q,i-1) \Rightarrow
c^-(q,i)$ and since $U\models
\neg c^-(q,i)$, it follows that $U\models \neg c^-(q,i-1)$. Consequently, 
$U$ satisfies the formula $F_3(r,i-1)$ and, so, $U\models F_4(q,i-1)$.
It follows that $c(q,i-1)\in U$, a contradiction (we recall that $i_q=i$). 
Hence, there is $j$, $1\leq j\leq s$, such that $c(a_j,i-1)\in U$. It 
follows that $a_j\in [M(U)]_{i-1}$ and $[M(U)]_{i-1}\not=\emptyset$. 
\hfill$\Box$

\begin{lemma}\label{lem-13}
Let $U$ be a model of $T_0(P)$ and let $|M(U)|\leq k$. Then
\begin{enumerate}
\item $i_U\leq k$, and
\item $M(U)$ is a stable model of $P$. 
\end{enumerate}
\end{lemma}
Proof: (1) The assertion follows directly from the fact that $|M(U)|\leq
k$ and from Lemma \ref{lem-12}.\\
(2) We need to show that $M(U)=LM(P^{M(U)})$. We will first show that
${M(U)}\subseteq LM(P^{M(U)})$. Since ${M(U)}=\bigcup_{i=1}^{i_U} 
[M(U)]_i$, we will show 
that for every $i$, $1\leq i\leq i_U$, $[M(U)]_i\subseteq LM(P^{M(U)})$.
We will proceed by induction. 
Let $q\in [M(U)]_1$. It follows that there is a rule $r$ such that $U\models
F_3(r,1)$. Consequently, $r$ is of the form $r = q\leftarrow
\n(b_1),\ldots,\n(b_t)$ and $U \models \neg c(b_1)\wedge \ldots\wedge
\neg c(b_t)$. Hence, for every $j$, $1\leq j\leq t$, $b_j\notin M(U)$. 
Consequently, the rule $(q\leftarrow\ .)$ is in $P^{M(U)}$ and, so,
$q\in LM(P^{M(U)})$. The inductive step is based on a similar argument. It
relies on the inequality $i_U\leq k$ we proved in (1). We
leave the details of the inductive step to the reader. 
 
We will next show that $LM(P^{M(U)})\subseteq {M(U)}$. We will use the
characterization of $LM(P^{M(U)})$ as the limit of the sequence of iterations
of the van Emden-Kowalski operator $T_{P^{M(U)}}$: 
\[
LM(P^{M(U)})= \bigcup_{i=0}^{\infty}T_{P^{M(U)}}^i(\emptyset).
\] 
We will first show that for every integer $i$, $0\leq i\leq k+1$, we
have: $T^i_{P^{M(U)}}(\emptyset)\subseteq {M(U)}$ and for every 
$q\in T^i_{P^{M(U)}}(\emptyset)$, $i_q\leq i$.  

Clearly, $T^0_{P^{M(U)}}(\emptyset) = \emptyset \subseteq {M(U)}$. 
Hence, the basis 
for the induction is established. Assume that for some $i$, $0\leq i\leq
k$, $T^i_{P^{M(U)}}(\emptyset) \subseteq {M(U)}$ and that for every 
$q\in T^i_{P^{M(U)}}(\emptyset)$, $i_q\leq i$. Consider 
$q\in T^{i+1}_{P^{M(U)}}(\emptyset)$. If $U\models c^-(q,i+1)$, 
then $c(q,v)\in U$ for some $v$, $1\leq v\leq i$. Since $U\models
F_2(q)$, $c(q)\in U$ and $q\in {M(U)}$. By Lemma \ref{lem-11}, it follows
that $i_q = v$. Hence, $i_q < i+1$.

Thus, assume that $U\models \neg c^-(q,i+1)$. Since $q\in
T^{i+1}_{P^{M(U)}}(\emptyset)$, there is a rule 
\[
r = q\leftarrow a_1,\ldots,a_s,\n(b_1),
\ldots,\n(b_t)
\] 
in $P$ such that $b_j\notin {M(U)}$, for every $j$, 
$1\leq j\leq t$, and $a_j \in T^{i}_{P^{M(U)}}(\emptyset)$, $1\leq i\leq s$. 
By the induction hypothesis, for every $j$, $1\leq j\leq s$, we have
$a_j\in {M(U)}$ and $i_{a_j}\leq i$. It follows that $U\models F_3(r,i+1)$ and,
consequently, that $c(q,i+1)\in U$. Since $U\models F_2(q)$, $c(q)\in U$
and $q\in {M(U)}$. It also follows (Lemma \ref{lem-11}) that $i_q = i+1$.

Thus, we proved that $\bigcup_{i=0}^{k+1} T_{P^{M(U)}}^i(\emptyset)
\subseteq {M(U)}$. Since $|{M(U)}|\leq k$, there is $j$, $0\leq j\leq k$ such that
$T_{P^{M(U)}}^j(\emptyset)=T_{P^{M(U)}}^{j+1}(\emptyset)$. It follows that for
every $j'$, $j<j'$, $T_{P^{M(U)}}^j(\emptyset)=T_{P^{M(U)}}^{j'}(\emptyset)$.
Consequently, $T_{P^{M(U)}}^i(\emptyset) \subseteq {M(U)}$ for every non-negative
integer $i$. \hfill$\Box$

Consider now a stable model $M$ of the program $P$ and assume that 
$|M|\leq k$. Clearly, $M=\bigcup_{i=1}^\infty T^i_{P^M}(\emptyset)$. 
For each atom 
$q\in M$ define $s_q$ to be the least integer $s$ such that 
$q\in T^s_{P^M}(\emptyset)$. Clearly, $s_q\geq 1$. Moreover, since 
$|M|\leq k$, it follows that for each
$q\in M$, $s_q\leq k$. Now, define
\[
U_M = \{c(q), c(q,s_q) \colon q\in M\}\cup \{c^-(q,i)\colon q\in M,\
s_q<i\leq k+1\}
\]

\begin{lemma}\label{lem-14}
Let $M$ be a stable model of a logic program $P$ such that 
$|M|\leq k$. Under the terminology introduced above, the set of 
atoms $U_M$ is a model of $T_0(P)$.
\end{lemma}
Proof: Clearly, $U_M\models F_1(q,i)$ for $q\in \At(P)$ and 
$2\leq i\leq k+1$, and $U_M\models F_2(q)$ for $q\in \At(P)$. 

We will now show that $U_M\models F_4(q,i)$, for $q\in \At(P)$ and
$i=1,2,\ldots, k+1$. First, we will consider the case $q\in M$. There 
are three subcases here depending on the value of $i$. 

We start with $i$ such that $s_q < i \leq k+1$. Then $U_M\not\models 
\neg c^-(q,i)$. It follows that $U_M\not\models F_3(r,i)$ for every 
rule $r\in P$ such that $h(r)=q$.  Since $U_M\not \models c(q,i)$, 
$U_M\models F_4(q,i)$. 

Next, we assume that $i=s_q$. Then, there is a rule
$r = q\leftarrow a_1,\ldots,a_s,\n(b_1), \ldots,$ $\n(b_t)$
in $P$ such 
that $b_j\notin M$, for every $j$, $1\leq j\leq t$, and $a_j \in 
T^{i-1}_{P^M}(\emptyset)$, $1\leq j\leq s$.
Clearly, $U_M\models F_3(r,i)$. Since $U_M\models c(q,i)$, it follows that
$U_M\models F_4(q,i)$, for $i=s_q$. 

Finally, let us consider the case
$1\leq i <s_q$. Assume that there is rule $r\in P$ such that $h(r)=q$
and $U_M\models F_3(r,i)$. Let us assume that 
$r = q\leftarrow a_1,\ldots,a_s,\n(b_1), \ldots,$ $\n(b_t)$. 
It follows that for every $j$,
$1\leq j\leq t$, $U_M\models \neg c(b_j)$. Consequently, for every 
$j$, $1\leq j\leq t$, $b_j\notin M$ and the rule $r' = q\leftarrow
a_1,\ldots,a_s$ belongs to the reduct $P^M$. In addition, for every $j$,
$1\leq j\leq s$, $c^-(a_j,i)\in U_M$. Thus, $a_j\in M$ and $s_{a_j}\leq
i-1$. This latter property is equivalent to $a_j\in 
T^{i-1}_{P^M}(\emptyset)$. Thus, it follows that 
$q\in T^i_{P^M}(\emptyset)$ and $s_q\leq i$ --- a contradiction with
the assumption that $i<s_q$. Hence, for every rule $r$ with the head
$q$, $U_M\not\models F_3(r,i)$. Since for $i<s_q$, $c(q,i)\notin U_M$,
$U_M\models F_4(q,i)$.

To complete the proof, we still need to consider the case $q\notin M$. 
Clearly, for every $i$, $1\leq i\leq
k+1$, $U_M\not\models c(q,i)$. Assume that there is $i$, $1\leq i\leq
k+1$, and a rule $r$ such that $h(r)=q$ and $U_M\models F_3(r,i)$. Let
us assume that $r$ is of the form $q\leftarrow a_1,\ldots,a_s,\n(b_1), 
\ldots,\n(b_t)$. It follows that $c^-(a_j,i)\in U_M$ and, consequently,
$a_j\in M$ for every $j$, $1\leq j\leq s$. In addition, it follows 
that for every $j$, $1\leq j\leq t$, $U_M\models \neg c(b_j)$ and,
consequently, $b_j\notin M$. Thus, $q\leftarrow a_1,\ldots,a_s$ belongs 
to the reduct $P^M$ and, since $M$ is a model of the reduct, $q\in M$, a
contradiction. It follows that for every $i$, $1\leq i\leq k+1$, 
$U_M\models F_4(q,i)$.
\hfill$\Box$

For each atom $q\in \At(P)$, let us introduce $k^2+2k$ new atoms
$d(q,i)$, $1\leq i\leq k^2+2k$, and define
\[
T(P) = T_0(P) \cup \{c(q)\Leftrightarrow d(q,i)\colon 1\leq i\leq
k^2+2k\}.
\]
Lemmas \ref{lem-11} - \ref{lem-14} add up to a proof of the following
result.

\begin{theorem}\label{th-21}
Let $k$ be a non-negative integer and let $P$ be a logic program. 
The program $P$ has a stable model of size at most $k$ if and only if 
the theory $T(P)$ has a model of size at most $(k+1)(k^2+2k)$.
\end{theorem}
Proof: $(\Rightarrow)$ Let $M$ be a stable model of $P$ such that
$|M|\leq k$. By Lemma \ref{lem-14}, the set $U_M$ is a model of $T_0(P)$
Consequently, the set
\[
U = U_M\cup \{d(q,i)\colon q\in M, 1\leq i\leq k^2+2k\}
\]
is a model of $T(P)$. Moreover, it is easy to see that $|U_M|\leq 
2k+k^2$. Hence, $|U|\leq 2k+k^2 +k(k^2+2k)=(k+1)(k^2+2k)$. 

Conversely, let us assume that some set $V$, consisting of atoms
appearing in $T(P)$ and such that $|V|\leq (k+1)(k^2+2k)$, is a model of 
$T(P)$. Let us define $U$ to consist of all atoms of the form $c(q)$,
$c(q,i)$ and $c^-(q,i)$ that appear in $V$. Clearly, $U$ is a model of
$T_0(P)$. Let us assume that $M(U) \geq k+1$ (we recall that the
notation $M(U)$ was introduced just before Lemma \ref{lem-11} was stated). 
Then, there are at least $(k+1)(k^2+2k)$ atoms of type $d(q,i)$ in $V$. 
Consequently, $V > (k+1)(k^2+2k)$ as it contains also at least $k+1$ 
atoms $c(q)$, where $q\in M(U)$. This is a contradiction. Thus, it follows 
that $|M(U)|\leq k$. Moreover, 
by Lemma \ref{lem-13}, $M(U)$ is a stable model of $P$. \hfill$\Box$

Let us now define the following sets of formulas. First, for each atom 
$q\in \At(P)$ we define
\[
C_0(q) = \{\neg c(q)\vee d(q,i)\colon 1\leq i\leq k^2+2k\} \cup
         \{c(q)\vee \neg d(q,i)\colon 1\leq i\leq k^2+2k\}.
\]
Next, we define
\[
C_1(q,i) = \{\neg c^-(q,i) \vee c(q,1)\vee \ldots \vee c(q,i-1)\} \cup
           \{\neg c(q,j)\vee c^-(q,i)\colon 1\leq j\leq i-1\},
\] 
\[
C_2(q) = \{\neg c(q) \vee c(q,1)\vee \ldots \vee c(q,k+1)\} \cup
           \{\neg c(q,j)\vee c(q)\colon 1\leq j\leq k+1\},
\]
and
\[
C_4(q,i) = \{\neg c(q,i) \vee F_3(r_1,i)\vee \ldots \vee F_3(r_v,i)\} \cup
           \{\neg F_3(r_j,i)\vee c(q,i)\colon 1\leq j\leq v\},
\]
where $\{r_1,\ldots,r_v\}$ is the set of all rules in $P$ with $q$ in the
head.  

Clearly, the theory
\begin{eqnarray*}
T^c(P) &=&\{C_0(q)\colon q\in At(P)  \}\cup
     \{C_1(q,i)\colon q\in At(P), 2\leq i\leq k+1  \}\cup \\
     & &\{C_2(q)\colon q\in At(P)  \}\cup 
     \{C_4(q,i)\colon q\in At(P), 1\leq i\leq k+1 \}
\end{eqnarray*}
is equivalent to the theory $T(P)$. Moreover, it is a collection of sums
of products of literals. Therefore, it is a 3-normalized formula.
By Theorem \ref{th-21}, it follows that the problem $\SSM$ can be reduced to 
the problem $WS^\leq(3)$. Thus, we get the following result.

\begin{theorem}
The problem $\SSM(k)\in W[3]$.
\end{theorem} 

Next, we will show that the problem $WS^\leq(2)$ can be reduced to the
problem $\SSM$. Let $C =\{c_1,\ldots,c_p\}$ be a collection of
clauses. Let $A=\{x_1,\ldots,x_r\}$ be the set of atoms appearing in
clauses in $C$. For each atom $x\in A$, introduce $k$ new atoms
$x(i)$, $1\leq i\leq k$. By $S_i$, $1\leq i\leq k$, we denote the logic
program consisting of the following $n$ clauses:
\begin{center}
$x_1(i)\leftarrow \n(x_2(i)),\ldots, \n(x_r(i))$\\
$\cdots$\\
$x_r(i)\leftarrow \n(x_1(i)),\ldots, \n(x_{r-1}(i))$
\end{center}
Define $S = \bigcup_{i=1}^k S_i$. Clearly, each stable model of $S$ is
of the form $\{x_{j_1}(1),\ldots,x_{j_k}(k)\}$, where $1\leq j_p\leq r$
for $p=1,\ldots, k$. Sets of this form can be viewed as representations
of nonempty subsets of the set $A$ that have no more than $k$ elements.
This representation is not one-to-one, that is, some subsets have
multiple representations.

Next, define $P_1$ to be the program consisting of the clauses
\[
x_j\leftarrow x_j(i),\ \ \ j=1,\ldots,r,\ \ i=1,2,\ldots, k.
\]
Stable models of the program $S\cup P_1$ are of the form
$\{x_{j_1}(1),\ldots,x_{j_k}(k)\}\cup M$, where $M$ is a nonempty subset
of $A$ such that $|M|\leq k$ and $x_{j_1},\ldots,x_{j_k}$ enumerate
(possibly with repetitions) all elements of $M$. 

Finally, for each clause 
\[
c = a_1\vee \ldots \vee a_s\vee \neg b_1\vee \ldots \vee \neg b_t
\]
from $C$ define a logic program clause $p(c)$:
\[
p(c) = f\leftarrow b_1,\ldots,b_t,\n(a_1),\ldots,\n(a_s), \n(f)
\]
where $f$ is yet another new atom. Define $P_2=\{p(c)\colon c\in C\}$
and $P^C = S\cup P_1\cup P_2$.

\begin{theorem}
A set of clauses $C$ has a nonempty model with no more than $k$ elements
if and only if the program $P^C$ has a stable model with no more than
$2k$ elements.
\end{theorem}
Proof: Let $M$ be a nonempty model of $C$ such that $|M|\leq k$. Let
$x_{j_1},\ldots,x_{j_k}$ be an enumeration of all elements of $M$
(possibly with repetitions). Then the set $M'=\{x_{j_1}(1),\ldots, 
x_{j_k}(k)\}\cup M$ is a stable model of the program $S\cup P_1$. Since 
$M$ is a model of $C$, it follows that  
$(P^C)^{M'} = (S\cup P_1)^{M'} \cup F$, where $F$ consists of the clauses
of the form
\[
f\leftarrow b_1,\ldots,b_t,
\]
such that $t\geq 1$ and for some $j$, $1\leq j\leq t$, $b_j\notin M'$.
Since $M' = LM((S\cup P_1)^{M'})$, it follows that
\[
M' = LM((S\cup P_1)^{M'} \cup F) = LM((P^C)^{M'}).
\]
Thus, $M'$ is a stable model of $P^C$. Since $|M'|\leq 2k$, the ``only if'' 
part of the assertion follows.

Conversely, assume that $M'$ is a stable model of $P^C$. Clearly,
$f\notin M'$. Consequently, 
\[
LM((S\cup P_1)^{M'}) = LM((S\cup P_1\cup P_2)^{M'}) = LM((P^C)^{M'}) = M'.
\]
That is, $M'$ is a stable model of $S\cup P_1$. As mentioned earlier, it 
follows that $M' = \{x_{j_1}(1),\ldots, x_{j_k}(k)\}\cup M$, where $M$ is a
nonempty subset of $\At(P)$ such that $|M|\leq k$ and
$x_{j_1},\ldots,x_{j_k}$ is an enumeration of all elements of $M$.  

Consider a clause $c = a_1\vee \ldots \vee a_s\vee \neg b_1\vee \ldots
\vee \neg b_t$ from $C$. Since $M'$ is a stable model of $P^C$, it is a
model of $P^C$. In particular, $M'$ is a model of $p(c)$. Since $f\notin
M'$, it follows that $M'\models c$ and, consequently, $M\models c$.
Hence, $M$ is a model of $C$. \hfill $\Box$

Now the reducibility of the problem $WS^\leq(2)$ to the problem $\SSM$ is
evident. Given a collection of clauses $C$, to check whether it has a
model of size at most $k$, we first check whether the empty set of atoms
is a model of $C$. If so, we return the answer YES and terminate the
algorithm. Otherwise, we construct the program $P^C$ and check whether
it has a stable model of size at most $2k$. Consequently, we obtain the
following result.

\begin{theorem}
The problem $\SSM$ is {\rm W[2]}-hard.
\end{theorem}

\section{Open problems and conclusions}

The paper established several results pertaining to the problem of
computing small and large stable models. It also brings up interesting
research questions.

First, we proved that the problem $\LSM$ is in the class FPT. For
problems that are fixed-parameter tractable, it is often possible to
design an algorithm running in time $O(p(N)+f(k))$, where $N$ is the
size of the problem, $k$ is a parameter, $p$ is a polynomial and $f$ 
is a function \cite{df97}. Such algorithms are often practical for quite
large ranges of $N$ and $k$. The algorithm for the $\LSM$ problem
presented in this paper runs in time $O(m2^{k+k^2})$. It seems plausible
it can be improved to run in time $O(m+f(k))$, for some function $f$.
Such an algorithm would most certainly be practical for wide range of
values of $m$ and $k$. We propose as an open problem the challenge
of designing an algorithm for computing large stable models with this 
time complexity.

There is a natural variation on the problem of computing large
stable models: given a logic program $P$ and an integer $k$ (parameter),
decide whether $P$ has a stable model of size at least $|\At(P)|-k$.
This version of the problem $\LSM$ was recently proved by Zbigniew Lonc
and the author to be W[3]-hard (and, hence, fixed-parameter 
intractable) \cite{lt01}. The upper bound for the complexity of this
problem remains unknown.  

In the paper, we described an algorithm that for every fixed $k$,
decides the existence of stable models of size at most $k$ in time 
$O(n^{k-1}m)$, where $n$ is the number of atoms in 
the program and $m$ is its size. This algorithm offers only a slight 
improvement over the straightforward ``guess-and-check'' algorithm. 
An interesting and, it seems, difficult problem is
to significantly improve on this algorithm by lowering the exponent in
the complexity estimate to $\alpha k$, for some constant $\alpha< 1$.

We also studied the complexity of the problem $\SSM$ and showed that it
is fixed-parameter intractable. Our results show that $\SSM$ is
$W[2]$-hard. This result implies that the problem $\SSM$ is at least as hard 
as the problem to determine whether a CNF theory has a model of cardinality 
at most $k$, and strongly suggests that algorithms do not exist that
would decide problems $\SSM(k)$ and run in time $O(n^c)$, where $c$ is
a constant independent on $k$. For the upper bound, we proved in this 
paper that the problem $\SSM$ belongs to class $W[3]$. Recently, Zbigniew 
Lonc and the author \cite{lt01} showed that the problem $\SSM$ is, in 
fact, in the class $W[2]$.

\section*{Acknowledgments}
The author thanks Victor Marek and Jennifer Seitzer for useful 
discussions and comments. The author is grateful to anonymous referees 
for very careful reading of the manuscript. Their comments helped
eliminate some inaccuracies and improve the presentation of the results.
This research was supported by the NSF grants CDA-9502645, IRI-9619233 and 
EPS-9874764.


\end{document}